# Historical and Philosophical Reflections on the Einstein-de Sitter model

Cormac O'Raifeartaigh[a], Michael O' Keeffe[a] and Simon Mitton[b]

[a]*School of Science and Computing, Waterford Institute of Technology, Cork Road, Waterford, Ireland.*

[b]*St Edmund's College, University of Cambridge, Cambridge CB3 0BN, United Kingdom.*

Author for correspondence: coraifeartaigh@wit.ie

Abstract

We present some historical and philosophical reflections on the paper "*On the Relation Between the Expansion and the Mean Density of the Universe*", published by Albert Einstein and Willem de Sitter in 1932. In this famous work, Einstein and de Sitter considered a relativistic model of the expanding universe with both the cosmological constant and the curvature of space set to zero. Although the Einstein-deSitter model went on to serve as a standard model in 'big bang' cosmology for many years, we note that the authors do not explicitly consider the evolution of the cosmos in the paper. Indeed, the mathematics of the article are quite puzzling to modern eyes. We consider claims that the paper was neither original nor important; we find that, by providing the first specific analysis of the case of a dynamic cosmology without a cosmological constant or spatial curvature, the authors delivered a unique, simple model with a straightforward relation between cosmic expansion and the mean density of matter that set an important benchmark for both theorists and observers. We consider some philosophical aspects of the model and provide a brief review of its use as a standard 'big bang' model over much of the 20$^{th}$ century.

## 1. Introduction

In 1932, Albert Einstein and Willem de Sitter jointly proposed a relativistic model of the expanding universe in which the cosmological constant and the curvature of space were set to zero (Einstein and de Sitter 1932). This model, soon known as the Einstein-de Sitter universe, went on to become a standard model of modern 'big bang' cosmology. One reason was the model's great simplicity; by removing two major unknowns, the authors provided a cosmology with a simple relation between two observables, the rate of spatial expansion and the mean density of matter, that could be tested against astronomical observation. Another reason was that the critical density of matter characteristic of the model provided a useful benchmark for the classification of cosmic models by theorists. In any event, the theory became the prototype 'big bang' model for much of the 20$^{th}$ century (Nussbaumer and Bieri 2009 p. 152; Realdi 2019). For example, in the long debate between steady-state and 'big bang' cosmologies in the 1950s and 60s, the Einstein-de Sitter model became the default example of the latter (Kragh 2007, pp. 215-216; Kragh 2014). In the 1970s and 80s, the long search by astronomers to establish a reliable estimate of the mean density of matter was conducted with reference to the Einstein-de Sitter model, although it became increasingly clear that the density lay far below the critical value required for flatness (Longair 2006 pp. 340-361; Peebles 2020 pp. 105-111). With the hypothesis of cosmic inflation in the 1980s, the model returned to the fore, at least for many theoreticians, and by the early 2000s, precision measurements of the cosmic microwave background had provided the first observational evidence of a universe of Euclidean geometry. However, by this time, evidence had emerged of a dark energy component for the energy density of the universe, giving rise to today's Lambda-Cold-Dark-Matter or ΛCDM model (Martínez and Trimble 2009; Calder and Lahav 2010).

It is therefore quite surprising for the historian to note that the cosmic model proposed by Einstein and de Sitter in 1932 is not, strictly speaking, a model of the 'big bang' type, in the sense that the authors do not explicitly consider the evolution of the cosmos in their paper. On reflection, this should not come as too great a surprise. While the hypothesis of a universe expanding outwards from an initial singularity was first considered by Alexander Friedman in 1922 (Friedman 1922), this work did not receive much attention until the 1930s (Nussbaumer and Bieri 2009 pp. 110-11; Kragh 2014). On the other hand, the notion of an expanding cosmos with a physical 'fireworks' origin was first mooted by Lemaître in 1931 (Lemaître 1931b, 1931c); however, this model was considered highly speculative by most theorists for many years (Kragh 2008). It is thus interesting to the historian of science that a paper that later served as an archetypal 'big bang' model may not, as originally formulated, have been seen by the authors in this light.[1]

With this is mind, we thought it useful to provide a guided tour and analysis of the 1932 paper by Einstein and de Sitter. The astute reader will soon notice that the authors make use of the quantities $R_A$, the radius of Einstein's static universe of 1917 (Einstein 1917), and $R_B$, a characteristic length from de Sitter's empty cosmology of 1917 (de Sitter 1917). Indeed, in some ways the paper can be read as a closure of the decade-long debate between Einstein and de Sitter concerning the relative merits of their respective models of 1917.[2] After the tour, we discuss some further historical and philosophical aspects of the paper and in section 5, we provide a brief history of the model as the prototype 'big bang' model over much of the 20th century.

2. **Historical context**

With the publication of Edwin Hubble's observation of an approximately linear relation between the redshifts of the spiral nebulae and their distance (Hubble 1929), many theorists began to consider the possibility of a universe of expanding radius. Thus, the Einstein-de Sitter paper should be considered in the context of a number of works on relativistic cosmology that were published in the early 1930s by leading theorists such as Georges Lemaître, Arthur Stanley Eddington, Willem de Sitter, Albert Einstein, Howard Percy Robertson and Richard Tolman. The earliest of these cosmologies followed Lemaître's lead in proposing an 'emergent' universe that expanded from a large, pre-existing static radius (Lemaître 1927, 1931a; Eddington 1930, 1931; de Sitter 1930a, 1930b). Indeed, Eddington's consideration of the expansion of the universe as a result of instabilities in the static Einstein universe led to such models becoming known as Eddington-Lemaître models (Bondi 1952 pp. 84, 118; North 1965 pp. 122-125). However, as this work progressed, attention also turned to other cosmologies, in particular models that expanded from a singularity in the manner first suggested by Alexander Friedman (Friedman 1922; de Sitter 1930c, 1931a; Einstein 1931; Robertson 1932; Tolman 1930, 1932). In all of these early models of the expanding universe, there was little consideration of the evolution the cosmos in the distant past; the emphasis was on whether relativity could account for present astronomical observations, i.e., the redshifts of the nebulae (North 1965 pp. 125-126). We note also that almost all of these models assumed closed spatial curvature, in the same manner as the static cosmologies of Einstein and de Sitter (Einstein 1917; de Sitter 1917) and the early dynamic cosmologies of Friedman and Lemaître (Friedman 1922; Lemaître 1927).

---

[1] Despite hundreds of references to the paper, we are aware of only one article that notes that the Einstein-de Sitter paper is not a 'big bang' model (Kragh 1997).
[2] See (Kerzberg 1989; Realdi and Peruzzi 2009) for a discussion of this debate.

In particular, Einstein lost little time in investigating whether relativity could account for a non-static cosmos without the use of the cosmological constant term in the field equations. Adopting Alexander Friedman's analysis of a relativistic universe of positive spatial curvature expanding outwards from a singularity (Friedman 1922), Einstein set the cosmological constant to zero and arrived at a model of a universe that first expands and then contracts, a cosmology that is sometimes known as the Friedman-Einstein universe (Einstein 1931). Meanwhile, Willem de Sitter pursued a much more general investigation of expanding models (de Sitter 1930a, 1930b, 1930c, 1931a), although he appears to have retained a preference for emergent models (de Sitter 1931b, 1932).

It is known that the Einstein-de Sitter paper was written over the course of a few days in early January 1932, while Einstein and de Sitter were both visiting Caltech in Pasadena (figure 1). Indeed, the two great physicists were both housed at the famous Athenaeum at Caltech and worked intensely together for a few days (Eisinger 2011 p. 141; Guichelaar 2018 pp. 257-259). Given their decade-long debate on the relative merits of their first models of the cosmos, and the ground-breaking observations of astronomers at the nearby Mt Wilson Observatory, it is no surprise that their conversation turned to models of the expanding universe. Indeed, it is known that de Sitter also had many discussions with the Mt Wilson astronomers Edwin Hubble and Milton Humason during this period, concerning their ongoing observations of the redshift/distance relation of the spiral nebulae (Guichelaar 2018 pp. 257-259). During this time, de Sitter also gave a series of lectures at Caltech on relativity, cosmology and the expanding universe. It is known that Einstein attended at least one of these lectures and commented most favourably on it (Guichelaar 2018 p. 257). Of course, Einstein himself was no stranger to cosmological inspiration in sunny Pasadena; his sojourn at Caltech the previous winter had inspired his cosmic model of 1931 (Eisinger 2011 pp. 109-115).

We shall probably never know whether it was Einstein or de Sitter who drew the other's attention to Otto Heckmann's cosmological paper of 1931. In this article, Heckmann noted that a non-static cosmos could exhibit positive spatial curvature, negative curvature or no curvature at all (Heckmann 1931). Up to this point, almost every cosmic model had assumed positive spatial curvature.[3] Einstein and de Sitter immediately recognized that a dynamic, matter-filled universe of zero spatial curvature represented an intriguing class of cosmic models; with the cosmological constant and the pressure of matter set to zero, such a universe would expand indefinitely at an ever-slower rate of expansion. Most importantly, the model predicted a simple relation between two observables, the rate of expansion and the mean density of matter, that could be tested against observation. As regards the style of the paper, there is little question that the analysis is reminiscent of de Sitter's first models of the expanding universe, (de Sitter 1930a,1930b, 1930c), with little consideration of the origins and evolution of the universe.

---
[3] This point is sometimes disputed and will be discussed further in section 4.

### 3. A guided tour of the paper

Einstein and de Sitter begin their paper by noting an observation by the German theorist Otto Heckmann that, in a matter-filled universe of dynamic radius, positive spatial curvature is not a given:

> In a recent note in the Göttinger Nachrichten, Dr. O. Heckmann has
> pointed out that the non-static solutions of the field equations of the
> general theory of relativity with constant density do not necessarily
> imply a positive curvature of three-dimensional space, but that this
> curvature may also be negative or zero.

Although an exact reference is not given, there is little question that the authors are referring to Heckmann's 1931 paper *'Über die Metrik des sich ausdehnenden Universums'* (Heckmann 1931) or *'On the Metric of the Expanding Universe'*. In this paper, Heckmann points out that an expanding universe may be of positive, negative or zero spatial curvature. However, he does not explicitly explore the case of flat geometry, as discussed in section 4.

In the second paragraph of their paper, Einstein and de Sitter point out that neither the sign nor the magnitude of spatial curvature can be determined from current observations. An interesting question arises, namely, whether a cosmic model entirely devoid of spatial curvature can account for observations such as the rate of expansion and the density of matter:

> There is no direct observational evidence for the curvature, the only
> directly observed data being the mean density and the expansion, which
> latter proves that the actual universe corresponds to the non-statical
> case. It is therefore clear that from the direct data of observation we
> can derive neither the sign nor the value of the curvature, and the question
> arises whether it is possible to represent the observed facts without
> introducing a curvature at all.

In the third paragraph, the authors recall that the cosmological constant term was introduced to the field equations in order to account for a finite density of matter in a universe that was assumed to be static. For the case of a non-static universe, this term may not be necessary:

> Historically the term containing the "cosmological constant" λ was introduced into the field equations in order to enable us to account theoretically for the existence of a finite mean density in a static universe. It now appears that in the dynamical case this end can be reached without the introduction of λ.

The authors then give an extremely short passage of relativistic analysis. Assuming a time-dependent line element with no spatial curvature, and setting both the cosmological constant and the pressure of matter to zero, a differential equation[4] can be derived from the field equations that relates the fractional expansion of cosmic radius with the density of matter $\rho$:

> If we suppose the curvature to be zero, the line-element is
> $$ds^2 = -R^2 (dx^2\, dy^2 + dz^2) + c^2\, dt^2 \qquad (1)$$
> where $R$ is a function of $t$ only, and $c$ is the velocity of light. If, for the sake of simplicity, we neglect the pressure $p$, the field equations without λ lead to two differential equations, of which we need only one, which in the case of zero curvature reduces to:
> $$\frac{1}{R^2}\left(\frac{dR}{cdt}\right)^2 = \frac{1}{3}\kappa\rho \qquad (2)$$

In the fifth paragraph, the authors note that the fractional rate of expansion and the density of matter can be derived from observation, and relate these parameters to the characteristic lengths $R_A$ and $R_B$:

> The observations give the coefficient of expansion and the mean density:
> $$\frac{1}{R}\frac{dR}{cdt} = h = \frac{1}{R_B}\;;\quad \rho = \frac{2}{\kappa R_A^{\,2}} \qquad (2')$$

In the first expression of (2'), the authors are preparing to use Hubble's measurements of the redshift/distance relation of the nebulae as an empirical estimate of the fractional rate of cosmic expansion; as defined, the quantity $h$ is simply the Hubble constant divided by the speed of light. The quantity $R_B$ is not explicitly described in the text of the paper; from equation (2')[5] and from previous papers by de Sitter (de Sitter 1917, 1930a, 1930b), it corresponds to the radius[6] of the de Sitter universe (this model was known to the authors for many years as solution *B*). Similarly, in the second relation of equation (2'), the density of matter is defined relative to a quantity $R_A$ that is not specifically described in the text; from the equation presented, it corresponds to the radius of the static Einstein universe, first proposed in 1917 (Einstein 1917) and known to the authors as solution *A*.

---

[4] Equation (2) is a special case of the so-called Friedman equation, although the authors don't state this. The Einstein constant $\kappa$ is given by $\kappa = 8\pi G/c^2$

[5] As this equation is not numbered in the original paper, we use the label (2') for reasons of clarity.

Einstein and de Sitter then proceed to put theory together with observation. Taking a value of 500 km s$^{-1}$ Mpc$^{-1}$ for the Hubble constant, they first compute a value for $R_B$ using equation (2'). Since equations (2) and (2') imply the relation $R_A{}^2/R_B{}^2 = 2/3$, they then calculate a value for $R_A$ and from this the matter density is estimated:

> Therefore we have, from (2), the theoretical relation
> $$h^2 = \frac{1}{3}\kappa\rho \qquad (3)$$
> *or*
> $$\frac{R_A{}^2}{R_B{}^2} = \frac{2}{3} \qquad (3')$$
> Taking for the coefficient of expansion
> $$h = 500 \text{ km./sec. per } 10^6 \text{ parsecs}, \qquad (4)$$
> or
> $$R_B = 2 \times 10^{27} \text{ cm.},$$
> we find
> $$R_A = 1.63 \times 10^{27} \text{ cm.},$$
> or
> $$\rho = 4 \times 10^{-28} \text{ gr. cm}^{-3}, \qquad (5)$$
> which happens to coincide exactly with the upper limit for the density adopted by one of us.

We note first a slight inconsistency in notation and units. The "co-efficient of expansion" $h$ is defined in equation (2') as $h = (1/R).(dR/cdt)$ and has the dimensions of inverse length. On the other hand, the observational parameter $h = 500$ km s$^{-1}$ Mpc$^{-1}$ cited in equation (4) has the units of inverse time; this latter quantity is usually denoted as $H_0$ and really corresponds[7] to $hc$. More importantly, we note that the authors could have calculated the density of matter directly from the observed rate of expansion with the use of equation (3); taking $h = H_0/c$ and the Einstein constant $\kappa$ as 1.866 x 10$^{-26}$ m/kg, we obtain $\rho = 3h^2/\kappa \sim 4 \times 10^{-25}$ kg/m$^3$. Instead, the authors calculate the matter density by means of the ratio of the squares of the characteristic lengths $R_A$ and $R_B$. It is not clear why the authors chose this circuitous route, but the method owes much to de Sitter's first models of the expanding universe (de Sitter 1930a, 1930c, 1931), and may be an indication that the analysis is more of a sketch rather than a rigorous model. The result is of course the same, and the authors are pleased to note that their estimate for the density of matter is not inconsistent with values estimated by de Sitter from astronomical observations (de Sitter 1931a).

---

[6] We use the word 'radius' in a loose sense here, following de Sitter.
[7] The underlying reason for this is that the equation relating the Doppler shifts of the nebulae to cosmic expansion is given by $R'/R = v/cr$ where *r* is the distance of the source (Lemaître 1927).

In the last section of the paper, the authors consider the uncertainty in observational estimates of the rate of cosmic expansion and of the matter density. They note that the main source of error in determining each of these parameters lies in the significant uncertainty associated with the distances of the nebulae. Errors in observational estimates of the matter density may also arise due to the assumption that all of the material mass of the universe resides in the nebulae, although the authors doubt this assumption will introduce any appreciable error:

> The determination of the coefficient of expansion $h$ depends on the measured red-shifts, which do not introduce any appreciable uncertainty, and the distances of the extra-galactic nebulae, which are still very uncertain. The density depends on the assumed masses of these nebulae and on the scale of distance, and involves, moreover, the assumption that all the material mass in the universe is concentrated in the nebulae. It does not seem probable that this latter assumption will introduce any appreciable factor of uncertainty.

The authors then consider the ratio of the observables $h^2$ and $\rho$. Assuming a nebula occupying a spatial cube of side $1 \times 10^6$ lightyears, they note that their derived density of $4 \times 10^{-28}$ g/cm$^3$ corresponds to a mass of $2 \times 10^{11}$ solar masses for the galaxy. This is a second check on their estimate of matter density, as this estimate of galactic mass is not inconsistent with estimates from astronomy:

> Admitting it, the ratio $h^2/\rho$, or $R_A^2/R_B^2$, as derived from observations, becomes proportional to $\Delta/M$, $\Delta$ being the side of a cube containing on the average one nebula, and M the average mass of the nebulae. The values adopted above would correspond to $\Delta = 10^6$ light years, M = $2 \times 10^{11}$ ☉, which is about Dr. Oort's estimate of the mass of our own galactic system.

Thus, the authors conclude that a cosmic model that assumes no spatial curvature gives an estimate for the density of matter that is not inconsistent with observation:

> Although, therefore, the density (5) corresponding to the assumption of zero curvature and to the coefficient of expansion (4) may perhaps be on the high side, it certainly is of the correct order of magnitude, and we must conclude that at the present time it is possible to represent the facts without assuming a curvature of three dimensional space.

Finally, the authors stress that the spatial curvature may not in fact be zero, and suggest that an increase in the precision of observational data will allow for the determination of its sign and value:

> The curvature is, however, essentially determinable, and an increase in the precision of the data derived from observations will enable us in the future to fix its sign and to determine its value.

## 4. Discussion

*4.1 On spatial curvature and the cosmological constant*

Considering the cosmological constant first, we have noted above that Einstein had already removed this term in his 1931 model of the expanding cosmos (Einstein 1931). Although this paper is far less well-known than the 1932 paper of Einstein and de Sitter, it offers many insights into Einstein's cosmology.[8] We note here that Einstein's justification for the removal of the cosmological constant term in his 1931 model is identical to that of the present paper, namely that one could account for a finite density of matter in an expanding universe without it (section 3). We also note that in the 1931 model, Einstein derived a relation between the rate of expansion and the mean density of matter that is mathematically very similar to that of the present paper. However, an important difference is that the 1931 derivation necessitated several assumptions and approximations concerning the current phase of the cosmos in its timeline of evolution that are obviated in the 1932 paper by setting the spatial curvature to zero (O'Raifeartaigh and McCann 2014).

It is also true that the 1932 paper by Einstein and de Sitter was not the first in which it was noted that non-static cosmologies allow the possibility of a universe with no spatial curvature. Many commentators have suggested that, quite apart from Otto Heckmann, the possibility of Euclidean geometry for the cosmos had been explored by Alexander Friedman, Georges Lemaître and Howard Percy Robertson. On this basis, it has often been suggested that the Einstein-de Sitter paper of 1932 was neither original nor significant (Kragh 2007 p. 156; Nussbaumer 2014) and it has even been suggested that the paper would hardly have been published had it been submitted by less illustrious authors (Nussbaumer and Bieri 2009 p. 150; Barrow 2011 p. 75).

We do not agree with this view. Considering the case of Friedman first, there is little question that he delivered a comprehensive analysis of static and non-static cosmologies of positive spatial curvature in 1922 (Friedman 1922) and an analysis of static and non-static cosmologies of negative curvature in 1924 (Friedman 1924). However, we find no evidence that Friedman explored the specific case of a universe of flat geometry in any of his major publications. Turning to the case of Lemaître, it is certainly true that, in his 1925 analysis of the de Sitter model, Lemaître was led to the case of a time-varying universe of Euclidean geometry (Lemaître 1925). However, Lemaître did not analyse this cosmology, but dismissed it outright on the grounds that *"we are led... to the impossibility of filling up an infinite space with matter which cannot but be finite"* (Lemaître 1925). As for Robertson, it could be said that his 1929 exploration of a general (static or non-static) line-element for relativistic cosmology, based on general assumptions of homogeneity and isotropy, implicitly included the case of a universe of Euclidean geometry (Robertson 1929). However, this possibility is not explicitly explored and the physics of the paper is in any case firmly rooted in the context of a cosmos that is assumed to be static. Thus, we find that Einstein and de Sitter were correct to cite Heckmann as the first to consider the specific case of a time-varying universe of flat geometry, Even here, Heckmann touched on the case as one theoretical possibility amongst others and made no attempt to derive an expression for cosmic parameters that could be tested by observation (Heckmann 1931). By contrast, Einstein and de Sitter constructed a specific cosmic model with both spatial curvature and the cosmological constant set to zero with the express purpose of establishing a simple relation between the rate of expansion and the mean density of matter that could be compared with observation.

---

[8] We have given an analysis and first English translation of the paper in (O'Raifeartaigh and McCann 2014).

*4.2 On the rate of expansion and the density of matter.*

As noted above, setting both spatial curvature and the cosmological constant to zero in their model enabled the authors to derive a simple relation between the rate of expansion and the mean density of matter that could be tested against observation. Thus, taking Hubble's redshift/distance value of 500 km s$^{-1}$ Mpc$^{-1}$ for the fractional rate of expansion, they derived a value of $4 \times 10^{-28}$ g/cm$^3$ for the density of matter. (We have already noted that the authors do not obtain this estimate directly from the relation $\rho = 3h^2/\kappa$ (equation 3), but by a circuitous route involving the radii of the Einstein and de Sitter universes. They note that the resulting estimate for the mean density of matter lay at the upper bound of a range of values estimated by de Sitter from astronomical observations (de Sitter 1931a).

The authors provide a second check on their estimate of matter density by means of a simple order-of-magnitude calculation. Assuming the material mass of the universe is contained within the nebulae and assuming that a single nebula occupied a cubic volume of side $1 \times 10^6$ lightyears,[9] simple calculation suggested that the authors' estimate of the density of matter corresponded to a nebular mass of $2 \times 10^{11}$ solar masses. As they note, this estimate was consistent with estimates of the mass of the Milky Way provided by Jan Oort, de Sitter's colleague at the Leiden Observatory. Although a reference is not given, Oort's estimate was based on determinations of the local mass density of the Milky Way from stellar velocity dispersions and distributions that included a contribution from dark matter, in the tradition of earlier estimates by James Jeans and Jacobus Kapteyn (Oort 1932; Trimble 1990, 2013). Thus, it could be argued that the hypothesis of dark matter is implicit in the Einstein-de Sitter model (Longair 2004; Longair 2006 p. 116).

---

[9] This was a common assumption at the time (de Sitter 1930).

*4.3 On the evolution of the cosmos*

One of the most notable aspects of the Einstein-de Sitter paper is the lack of consideration of the evolution of the cosmos. Whereas relativistic cosmology leads naturally to the derivation of two independent differential equations from the field equations, the authors employed only one, equation (2) above. As pointed out in the fourth paragraph of the paper, only one differential equation is necessary to establish the main goal of the authors, a simple relation between the rate of spatial expansion and the density of matter that could be tested against observation. However, the omission confers a certain ambiguity on the model presented. While it is natural for the modern reader to assume a Friedman-type model in which the cosmos expands outwards from a singularity, this is not stated anywhere in the paper. Moreover, the use of the relation $\rho = 2/\kappa R_A^2$, characteristic of the static Einstein universe, is greatly puzzling, as is the manner in which the authors calculate the density of matter via the characteristic lengths $R_A$ and $R_B$. One possibility is that the authors have in mind a universe that expands outwards from a static radius, in the same manner as the first 'emergent' expanding models of de Sitter (de Sitter 1930a, 1931b). Another possibility is that the procedure is a mathematical artifice inherited from earlier models but no longer of physical significance. This interpretation would certainly be consistent with the impression of a model that was probably intended more as a rough sketch than as a rigorous analysis of the expanding cosmos.[10] After all, it is also true that the authors also fail to consider the stability of the model (McCoy 2020).

Whatever the meaning of the enigmatic quantity $R_A$, we note that Einstein had reformulated the model without reference to it within a year (Einstein 1933). Indeed, in his later reviews of cosmology, Einstein considered only Friedman-type models that expanded outwards from a point of apparently infinite density, attributing the puzzling timespan of such models to shortcomings of theory in describing the high-density conditions of the early universe (Einstein 1933, 1945 p. 128).

---

[10] A well-known anecdote from Eddington suggests that the authors themselves did not consider the work to be of great importance (Eddington 1940, p128; Nussbaumer and Bieri 2009, p152).

*4.4 On the philosophy of the model*

From a philosophical point of view, the innovative aspect of the Einstein-de Sitter paper was the proposal of a cosmic model of open spatial geometry. It is often forgotten that, following in the footsteps of Friedman and Lemaître, almost all[11] of the dynamic cosmic models that were proposed in years 1929-1932 in the wake of Hubble's observations were framed in terms of a cosmology of positive spatial curvature. The hypothesis of positive spatial curvature can be traced back to the very first relativistic model of the cosmos, Einstein's model of 1917 (Einstein 1917). Having struggled with the concept of a finite density of matter in an unbounded static cosmos, Einstein proposed a universe of closed, spherical curvature in order to satisfy his view of Mach's principle and the relativity of inertia (Einstein 1918; Barbour 1990; O'Raifeartaigh et al. 2017). That subsequent models followed this geometry is probably due to a loose expectation of the gravitational effect of matter. Einstein's expanding model of 1931 was of positive curvature and not in obvious conflict with his belief in Mach's principle; however, the same could hardly be said of the Einstein-de Sitter model of 1932. That Einstein proposed such a model is perhaps another indication of a gradual change in his attitude to Mach's principle in these years (Einstein 1949 p. 29; Tian Yu Cao 1997 pp. 93-94; Gutfreund and Renn 2017 p. 40).

We have noted earlier that Lemaître was led in his 1925 analysis of the de Sitter model to consider the case of a non-static cosmos of Euclidean geometry, but dismissed the possibility on the basis that infinite space could not be filled by a finite amount of matter. Here, it appears that Lemaître conflated the requirement of a finite mean density of matter in an expanding universe with a requirement of a finite quantity of matter. The most likely explanation for this uncharacteristic error is that, although Lemaître was well aware of the non-static character of the de Sitter metric in a mathematical sense, he was not truly thinking in 1925 in terms of a universe that is physically expanding.[12]

Thus, it could be argued that the Einstein-de Sitter paper of 1932 was an important advance in philosophical terms. The authors shook off the heritage of positive spatial curvature, a legacy from static models of the cosmos that had dogged the first tranche of expanding cosmologies. For the first time, physicists took seriously the prospect of an unbounded universe in which matter played a subsidiary role. More pragmatically, by setting the curvature to zero, along with the cosmological constant, the authors delivered a cosmic model with a simple relation between cosmic expansion and the density of matter that could be tested against observation.

---

[11] Friedman's paper of 1924 was not well-known.
[12] A similar observation can be made about the cosmologies of Hermann Weyl, Cornelius Lanczos and Howard P. Roberston (Nussbaumer and Bieri 2009 pp. 78-82).

It is interesting that, in contrast with Einstein's deliberations of 1917, few physicists appeared to have been concerned with the philosophical implications of a cosmos of open geometry. Yet such geometries were not without concern as, unlike models of closed curvature, they did not avoid the concept of an 'actual infinite' (North 1965 p. 135). As the theoretician Leopold Infeld later remarked (Infeld 1949 pp. 495-496):

> Yet every mathematician - if given the choice - would rather see our universe closed than open. There is mathematical beauty in such a universe which reveals itself when we consider any mathematical problem on such a cosmological background. In such a closed universe we have simple boundary conditions and we do not need to worry about infinities in time and space. Compared with the closed universe the open one of Einstein-de Sitter appears to be dulled and uninspired.

Some further considerations can be found by Lemaître in his 1929 work *La Grandeur de l'Espace* (Lemaître 1929; Lemaître 1950 pp. 22-56). However once again, he appears to conflate the issue of the counting of an infinite number of objects (stars and galaxies) with the postulate of a finite mean density of matter in infinite space. Eddington continued to argue strongly for a positive curvature of space (Eddington 1933 pp. 29-65). However, this argument was based on considerations of the dimensions of elementary particles and did not attract much support (Milne 1933 pp. 28-29; North 1965 pp. 281-282). In general, few theorists and astronomers seemed perturbed by the proposal of open geometry for the cosmos, perhaps an indication that they viewed the Einstein-de Sitter model as a useful hypothetical tool rather than a literal description of the universe. After all, the authors themselves suggest in the concluding section of the paper that spatial curvature of the cosmos could one day be observed: "*The curvature is, however, essentially determinable, and an increase in the precision of the data derived from observations will enable us in the future to fix its sign and to determine its value*" (Einstein and de Sitter 1932). It seems likely that most scholars saw little advantage in perusing the philosophical implications of the model until better estimates of cosmic parameters such as spatial curvature, material pressure and the cosmological constant were forthcoming from observation.[13]

## 5. The Einstein-de Sitter model as a standard model

The Einstein-de Sitter model became very well-known and went on to play a significant role in 20$^{th}$ century cosmology. For theorists, it marked an important hypothetical case in which the expansion of the universe was precisely balanced by a critical density of matter, given by equation (3) as $\rho_c(t) = 3h^2/\kappa = 3H^2(t)/8\pi G$. This allowed for a useful classification of cosmic models. Assuming a vanishing cosmological constant, a cosmos of mass density higher than the critical value would be of closed spatial geometry and eventually collapse, while a cosmos of mass density less than the critical value would be of open spatial geometry and expand at an ever increasing rate; in between lay the critical case of a cosmos with Euclidean geometry that would expand at an ever decreasing rate. Indeed, this classification of expanding models became a staple of cosmology textbooks (Bondi 1950 pp 82-86; Harrison 1981 p. 298). The geometry of such models was later usefully described in terms of the dimensionless density parameter $\Omega$, defined as the ratio of the actual matter density of the universe $\rho$ to the critical density $\rho_c$ required for spatial closure, i.e., $\Omega = \rho/\rho_c$. This simple classification scheme could be generalized to models with a cosmological constant and radiation pressure by defining the energy density parameter as $\Omega = \Omega_M + \Omega_\lambda + \Omega_R$, where $\Omega_M$, $\Omega_\lambda$ and $\Omega_R$ represented the energy density contributions due to matter, the cosmological constant and radiation respectively. In this scheme, the Einstein-de Sitter universe is neatly specified as ($\Omega$=1: $\Omega_M$ =1, $\Omega_\lambda$ =0, $\Omega_R$ =0) with $\Omega_M = (8\pi G/3H_0^2)\rho_M$. One immediately sees that the Einstein-de Sitter model represents a very special case as $\Omega = \Omega_M = 1$ for all time (Hawley and Holcomb 1998 pp 302-303; Liddle 1999 pp. 49-53).

The Einstein-de Sitter model also marked an important benchmark case for observers; in the absence of empirical evidence for spatial curvature or a cosmological constant, it seemed the cosmos could be described in terms of just two parameters, each of which could be determined by astronomy. In addition, it was soon realised that, in addition to astronomical methods such as galaxy counts and stellar dynamics, the mean density of matter could be estimated by measuring the rate of expansion at different epochs. Defining an expected slowing of the expansion over time in terms of a deacceleration parameter $q_0$, it was easily shown that for models without a cosmological constant $q_0 = \Omega_M/2$. One could therefore expect a value of $q_0 > 1/2$ for a cosmos of closed spatial curvature, $q_0 < 1/2$ for a cosmos of open geometry and $q_0 = 1/2$ for a universe of Euclidean geometry (Liddle 1999 pp. 50-53). Thus, the Einstein-de Sitter model soon became a significant benchmark model for astronomers (North 1965, p. 134; Kragh 1996, p. 35; Nussbaumer and Bieri 2009, p. 152).

---

[13] This could be labelled the 'shut up and observe' phase of cosmology in analogy with the well-known 'shut up and calculate' phase of quantum field theory (Kaiser 2011 pp. 1-25).

The opening of the 200-inch Hale telescope at the Palomar Observatory in California in 1949 heralded a new era of observational cosmology. In addition, the hypothesis of steady-state cosmology as an alternative to evolving models of the cosmos spurred new efforts to determine key cosmological parameters (Sandage 1961; Longair 2004). In this work, attention focused on the Einstein-de Sitter model and the determination of two parameters, the current rate of cosmic expansion $H_0$ and the deacceleration parameter $q_0$. Indeed, the challenge to establish observational values for these parameters was later dubbed "the search for two numbers" (Sandage 1970).[14]

By the mid-1960s, estimates of the rate of cosmic expansion had decreased by almost a factor of ten, temporarily easing the age problem associated with 'big bang' models (Longair 2006 pp. 340-361). By contrast, it became more and more apparent during the 1960s and 70s that estimates of the average density of matter from astronomy fell a long way below the critical value of the Einstein-de Sitter model. Even accounting for the existence of dark matter,[15] estimates of the matter density from galaxy counts and rotational dynamics remained below 30% of that required for flatness. Similarly, measurements of the deacceleration parameter $q_0$ also suggested a very low value for the density of matter (Sandage 1971; Longair 2006 pp. 340-361).

In the early 1980s, the theory of cosmic inflation was proposed in order to address numerous puzzles associated with evolving models of the universe (Guth 1981; Linde 1982; Albrecht and Steinhardt 1982). Inflation certainly gave new life to the prediction of a cosmos of Euclidean geometry, at least amongst many theorists, although others pointed out that estimates of the mean density of matter from astronomy remained far below the critical value (Coles and Ellis 1994; Peebles 2020 pp. 82-105). In the 1990s, new measurements of the Hubble constant from the Hubble Space Telescope, together with measurements of the cosmic microwave background, models of structure formation and constraints on the matter content of the universe set by primordial nucleosynthesis, caused a number of theorists to argue forcefully for a cosmic model with zero spatial curvature and a positive cosmological constant (Krauss and Turner 1995; Ostriker and Steinhardt 1995).[16] This proposal received an enormous boost when a new generation of observational programmes to measure the deacceleration parameter $q_0$ using supernovae as standard candles gave the first evidence of a *negative* deacceleration, i.e., of an acceleration in expansion (Riess et al. 1998; Perlmutter et al. 1999). In the early years of the 21st century, new precision measurements of the cosmic microwave background provided the first observational evidence that we do indeed inhabit a universe with spatial curvature close to zero. Put together, these measurements resulted in today's model of a universe of Euclidean geometry with energy contributions of $\Omega_M \sim 0.3$ and $\Omega_\Lambda \sim 0.7$ from matter and from dark energy respectively (Martínez and Trimble 2009; Calder and Lahav 2010).

---

[14] Note for example that steady-state models predicted a value of $q_0 = -1$.
[15] Strong observational evidence for the existence of dark matter emerged in the 1960s (Trimble 1990, 2013).

**Conclusions**

It is intriguing for the historian to note that, although the Einstein-de Sitter model of 1932 served as the prototype 'big bang' model for much of the 20th century, the authors do not explicitly consider the evolution of the cosmos in their paper. Indeed, a close study of the paper suggests a model that was probably intended as a rough sketch of the expanding cosmos. However, by providing the first analysis of the specific case of a dynamic cosmology without a cosmological constant or spatial curvature, the authors delivered a unique, simple model with a straightforward relation between cosmic expansion and the mean density of matter that became an important benchmark for both theorists and observers. The model also represented a significant philosophical advance, as it was the first well-known cosmology that was not of closed spatial curvature. While today's Λ-CDM model of a cosmos of Euclidean geometry with a positive cosmological constant is a much better fit to current observational data, the physical meaning of dark energy remains elusive.

**Acknowledgements**

Cormac O'Raifeartaigh thanks Professor Jim Peebles, Professor G. F. R. Ellis and Dr. Phillip Helbig for helpful discussions. Simon Mitton thanks St Edmund's College, University of Cambridge for the support of his research in the history of science.

---

[16] It should be noted that many theorists never truly dropped the cosmological constant (O'Raifeartaigh et al. 2018)

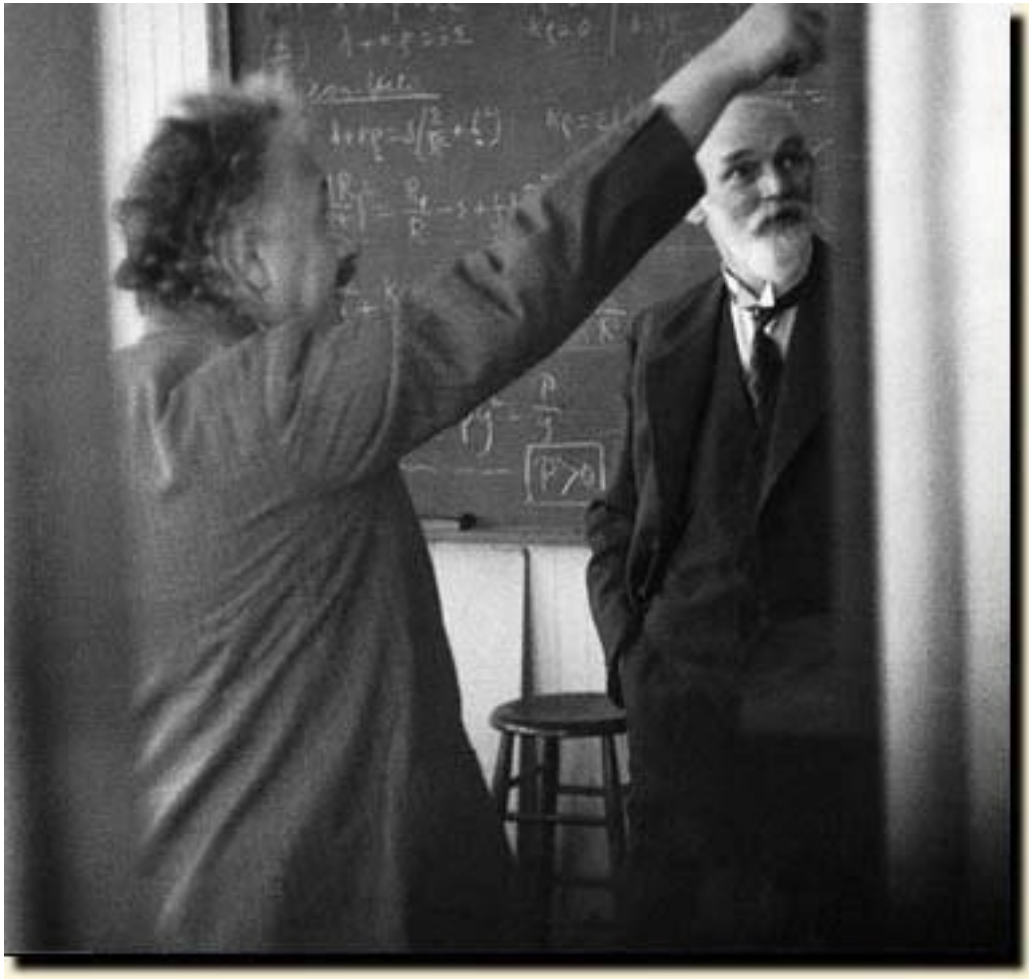

**Figure 1**

A famous photo of Einstein and de Sitter at work together at Caltech, Pasadena in January 1932. ©Associated Press.